\documentclass[conference]{IEEEtran}
\ifCLASSINFOpdf
\else
\fi
\usepackage[pdftex]{graphicx}
\usepackage{amsmath}

\usepackage[ruled, vlined]{algorithm2e}
\begin{document}
%
\title{Selectively Precoded Polar Codes}

\author{\IEEEauthorblockN{Samir Kumar Mishra and KwangChul Kim}
\IEEEauthorblockA{\textit{Device Solutions Research, Samsung Electronics} \\
Hwaseong-si, Gyeonggi-do, South Korea \\
\{samir.mishra,  marco.kim\}@samsung.com}
}


%


\maketitle

\begin{abstract}
In this paper, we propose \textit{selectively precoded polar (SPP) code}, built on top of Arikan's capacity achieving polar codes. We provide the encoding and decoding scheme for SPP code. Simulation results show that for a target frame erasure rate (FER) of $\mathbf{10^{-5}}$, a (128, 64) SPP code is just 0.23 dB away from the information theoretic limit at this blocklength. Further, it is also shown that such codes possess better distance properties compared to other contemporary polar code variants.
\end{abstract}


%
\IEEEpeerreviewmaketitle

\section{Introduction}
\IEEEPARstart{P}{olar} codes introduced by Arikan \cite{arikan2009channel} are the first provably capacity achieving codes for the class of binary input memoryless symmetric (BMS) channels with low encoding and decoding complexity of order $ O(N\log_2{N})$ for a code of blocklength $N$. Polar code is based on a phenomenon of channel polarization where a communication channel is transformed into polarized sub-channels: either completely noisy or noiseless. Information bits are transmitted over noiseless sub-channels, while fixed or \textit{frozen} bits are sent over the noisy ones. Polar codes are already being used in 5G New radio (NR) for encoding and decoding of control information.

\begin{figure}
\centering
\includegraphics[width=3.4in]{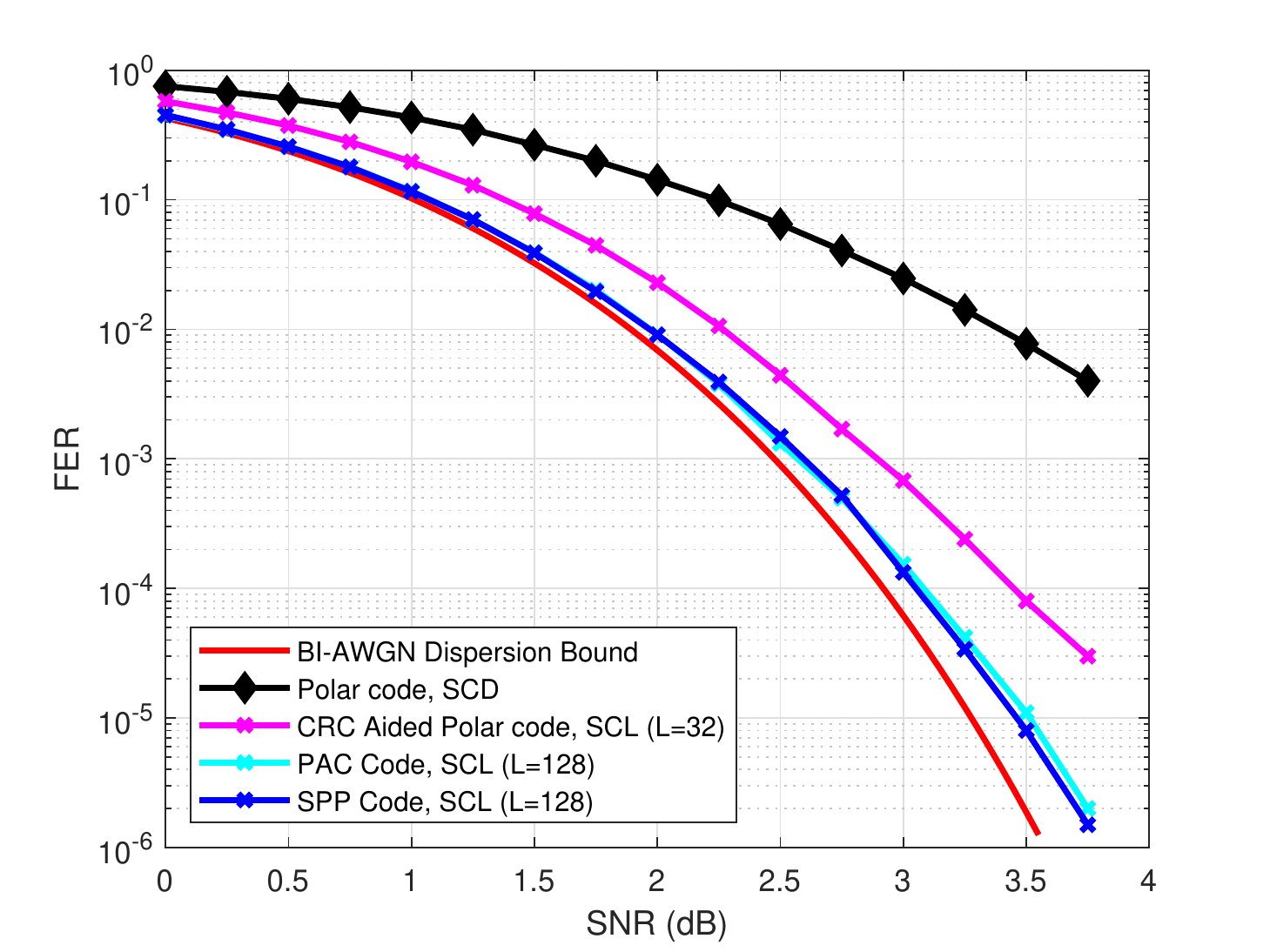}
\caption{FER Performance of (128, 64) polar code variants}
\label{fig1}
\end{figure}

Polar codes achieve channel capacity asymptotically as the blocklength $N$ of the code approaches infinity. However, for short blocklengths, the performance of polar codes is not good enough. Figure \ref{fig1} shows the performance of polar code and its variants for a blocklength $N=128$ and rate $R=0.5$ for a binary input additive white gaussian noise (BI-AWGN) channel. Figure \ref{fig1} also shows the BI-AWGN dispersion bound which is the minimum probability of error $\epsilon^{*}(N, R)$ that can be achieved on a BI-AWGN channel by using a code of blocklength $N$ and rate $R$ under maximum likelihood (ML) decoding. It can be observed clearly that there is a big gap between polar code with successive cancellation decoding (SCD) and the dispersion bound. This poor performance can be partly attributed to poor distance properties of polar codes and also the sub-optimality of SCD as compared to ML decoding \cite{tal2015list}.

Since Arikan's ground breaking work, there have been a lot of efforts to enhance the performance of polar code for short blocklengths, a survey of which can be found in \cite{cocskun2018efficient}. Specifically, CRC-Aided Polar codes under succesive cancellation list (SCL) decoding \cite{tal2015list} improve the performance quite a lot. Figure \ref{fig1} shows the FER performance of a (128, 72) polar code combined with a (72, 64) cyclic code which acts as CRC under SCL decoding with a list size $L=32$. This approach has been adopted to 5G NR standard and has remained the state of the art ever since.

In the Shannon Lecture at the International Symposium on Information Theory (ISIT) 2019, Arikan presented \textit{polarization assisted convolutional (PAC) codes}, which are significant improvement over the state-of-the-art polar codes \cite{arikan2019sequential}. Under sequential decoding, the FER performance of PAC codes is just 0.25 dB away from the BI-AWGN dispersion bound approximation at a target FER of $10^{-5}$. Further, it was observed in \cite{yao2020list} and \cite{rowshan2020polarization} that nearly same FER performance can as well be obtained by list decoding as shown in Figure \ref{fig1}.

In this paper, we propose \textit{selectively precoded polar (SPP) code}. SPP codes can be considered to a simple generalization of Arikan's PAC codes. As shown in Figure \ref{fig1}, the FER performance of SPP code is just 0.23 dB away from the BI-AWGN dispersion bound approximation at a target FER of $10^{-5}$. SPP codes have lesser encoding and decoding complexity compared to PAC codes.

The outline of this paper is as follows. In Section II, we first describe the coding scheme of SPP code. In Section III, we present numerical results and analyze them. In Section IV, we attempt to understand the good performance of SPP code by analyzing its weight distribution and comparing with other contemporary polar code variants. Finally, in Section V we conclude by mentioning some open problems.

\section{Selectively Precoded Polar Code}
\begin{figure}
\centering
\includegraphics[width=3.4in]{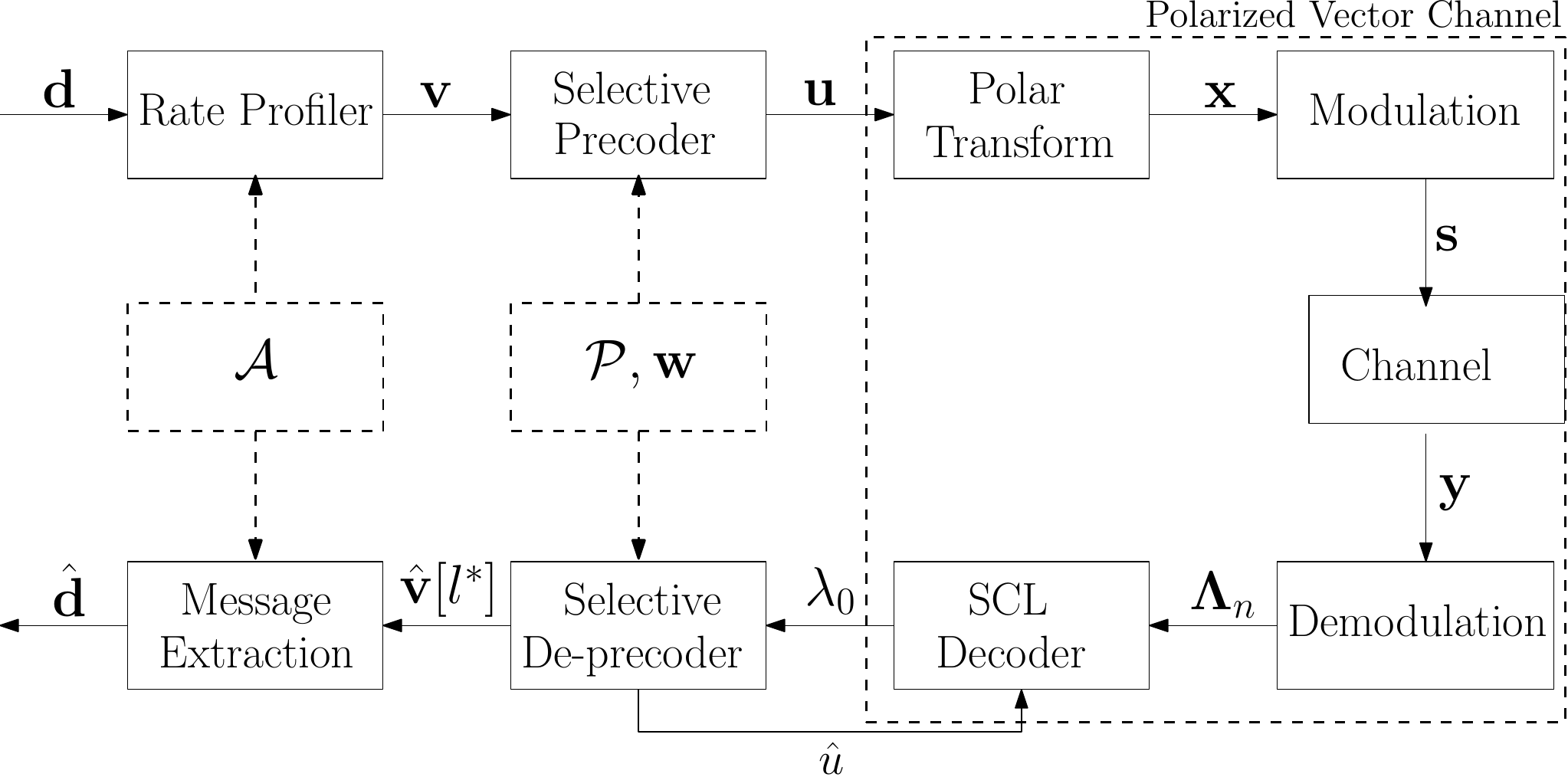}
\caption{Coding scheme of SPP code}
\label{fig2}
\end{figure}

The coding scheme of a SPP code is shown in Figure \ref{fig2}. In Figure \ref{fig2}, the solid blocks refer to actual blocks used in the communication system. The dotted blocks refer to the information provided to these blocks. A SPP code can be denoted as $SPP(N, K, \mathcal{A}, \mathcal{P}, \mathbf{w})$. Here, $K$ is the number of information bits. $N$ is the length of the codeword which is mostly a power of 2. $\mathcal{A} \subseteq \{ 0, 1, ..., (N-1)\}$ is the set of information bit indices. $\mathcal{P} \subseteq \{ 0, 1, ..., (N-1)\}$ is the set of indices where the bits need to be precoded. $\mathbf{w}$ is a precoding vector of length $p$ containing 0s and 1s. $R=K/N$ is the rate of the code. Using the polar code terminology, $\mathcal{F} = \mathcal{A}^{c}$ is the set of \textit{frozen} indices, where no information is transmitted. These indices are filled with zeros.

\subsection{Encoding scheme of SPP code}
The encoding scheme of SPP code is shown in Algorithm \ref{algo1} and can be divided into three steps which are described below.
\subsubsection{\textbf{Rate Profiling}} 
The term rate-profiling was coined by Arikan while describing PAC code \cite{arikan2019sequential}. A rate-profiler maps the vector of information bits denoted by $\mathbf{d} = [d_0, d_1, ..., d_{K-1}]$ to a vector of bits $\mathbf{v} = [v_0, v_1, ..., v_{N-1}]$ according to $\mathcal{A}$. In other words, the $K$ information bits in $\mathbf{d}$ are mapped to positions in $\mathbf{v}$ indicated by $\mathcal{A}$. The rest $(N-K)$ positions in $\mathbf{v}$ are filled with zeros.

Let us consider an example where $(N, K) = (8, 4)$, $\mathbf{d} = [d_0, d_1, d_2, d_3]$ and $\mathcal{A} = \{3, 5, 6, 7\}$. Then, $\mathbf{v} = [0, 0, 0, d_0, 0, d_1, d_2, d_3]$. 

Arikan mentioned two methods to design $\mathcal{A}$ which defines the rate-profiler \cite{arikan2019sequential}. First, the set $\mathcal{A}$ could be designed by any of the methods used for polar code construction, a good survey of which can be found in \cite{vangala2015comparative}. In this case, $\mathcal{A}$ denotes the indices of sub-channels in the polarized vector channel with high reliability. This approach is called \textbf{polar rate-profiling}. Second, a Reed-Muller (RM) approach could be used. Here, given $N$ indices $\{0, 1, ..., (N-1)\}$, we choose $K$ indices whose binary representation have the largest Hamming weights, with ties resolved arbitrarily. This approach is called \textbf{RM rate-profiling}.
\begin{algorithm}\label{algo1}
    \SetKwInOut{Input}{Input}
    \SetKwInOut{Output}{Output}

    \Input{$\mathbf{d}$, $\mathcal{A}$, $\mathcal{P}$, $\mathbf{w}$, $N$, $p$}
    \Output{$\mathbf{x}$}
    $\mathbf{v} \gets$ RateProfiler$(\mathbf{d}, \mathcal{A}, N)$ \\
    $\mathbf{u} \gets$ SelectivePrecoder$(\mathbf{v}, \mathcal{P}, \mathbf{w}, N, p)$ \\
    $\mathbf{x} \gets$ PolarTransform$(\mathbf{u})$ \\
    \Return $\mathbf{x}$; \\

    \SetKwFunction{FMain}{SelectivePrecoder}
    \SetKwProg{Fn}{subroutine}{:}{}
    \Fn{\FMain{$\mathbf{v}$, $\mathcal{P}$, $\mathbf{w}$, $N$, $p$}}{
    $\mathbf{u} \gets \mathbf{v}$ \\
    \For{$i = 0,1,\cdots,(N-1)$}{
      \If{$i\in\mathcal{P}$}
       {
         $u_{i} \gets$ PrecodeOneBit$(i, \mathbf{v}, \mathcal{P}, \mathbf{w}, p)$ \\
        }
      }
    \Return $\mathbf{u}$
    }
    
    \SetKwFunction{FMain}{PrecodeOneBit}
    \SetKwProg{Fn}{subroutine}{:}{}
    \Fn{\FMain{$i$, $\mathbf{v}$, $\mathcal{P}$, $\mathbf{w}$, $p$}}{
         $s \gets i$ \\
         $e \gets max(0, s-p+1)$ \\
         $\mathbf{f} \gets \mathbf{w}_{0}^{s-e}$ \\
         $\mathbf{a} \gets \mathbf{v}_{s}^{e}$ \\
         $u \gets \bigoplus(\mathbf{a}\mathbf{f}^{T})$ \tcp*[f]{Modulo-2 sum} \\
        \Return $u$
    }
    \caption{Encoding Algorithm of SPP Code}
\end{algorithm}

\subsubsection{\textbf{Selective Precoding}}
The second step of encoding is selective precoding which is shown as a subroutine in Algorithm \ref{algo1}. The inputs to the precoder are the output of rate-profiling $\mathbf{v}$, along with $\mathcal{P}$ and  $\mathbf{w}$. The key idea of selective precoding is as follows. The bit at index indicated by $\mathcal{P}$ is replaced by a linear combination of itself and the $(p-1)$ bits that precede it. The linear combination is decided by precoding vector $\mathbf{w}$. Precoding is a rate-1 operation because the code rate has already been satisfied by rate profiling in Step 1.

Let us continue the example stated earlier. Let $\mathbf{w} = [1, 1, 1]$, so, $p = 3$. Let $\mathcal{P} = \{0, 1, 2, 4\}$. Thus, after precoding, we get output $\mathbf{u} = [0, 0, 0, d_0, d_0, d_1, d_2, d_3]$.

\subsubsection{\textbf{Polar Transform}}
The final step of encoding is to pass the precoded vector $\mathbf{u}$ through a Polar Transform $\mathbf{P}_n$ to output encoded bit vector $\mathbf{x}$.
\begin{equation}
\label{eq1}
\mathbf{x} = \mathbf{u}\mathbf{P}_{n} = \mathbf{u}\mathbf{P}^{\bigotimes n}
\end{equation}
$\mathbf{P}_n$ is the $n^{th}$ Kronecker power of the basic Polar Transform $\mathbf{P}= \begin{bmatrix}
			1 & 0 \\
			1 & 1
			\end{bmatrix}$ proposed by Arikan in \cite{arikan2009channel}.
			
In the absence of precoding, $\mathbf{w} = [1]$ and SPP code falls back to polar code.

\subsection{Modulation, Transmission and Demodulation}
The vector $\mathbf{x}$ is suitably modulated to a symbol vector $\mathbf{s}$ and transmitted through the channel. Modulation can be done by any of the popular digital modulation schemes like M-ary phase shift keying (M-PSK) or M-ary Quadrature Amplitude Modulation (M-QAM). For example, in case of binary phase shift keying (BPSK), bit 0 is mapped to +1 and bit 1 is mapped to -1. The channel corrupts the transmitted symbols. Specifically, the AWGN channel adds gaussian noise $\mathbf{n}$ with distribution $\mathcal{N}(0, \sigma_{n}^{2})$ to the transmitted symbols.
\begin{equation}
\label{eq2}
\mathbf{y} = \mathbf{s} + \mathbf{n}
\end{equation}
The receiver gets noise corrupted symbols $\mathbf{y}$ from the channel. A suitable demodulator corresponding to the modulator is used to recover the transmitted symbols from $\mathbf{y}$. Preferably, a soft demodulator which can output soft decisions is used. A soft decision contains both value of the detected bit and information about its reliability. For BPSK, soft decision $\lambda_{i}$ for a received symbol $y_i$ can be generated in the form of a log-likelihood ratio (LLR) as shown below.
\begin{equation}
    \label{eq3}
    \lambda_{i} = \ln{\frac{Pr(y_i | s_i = +1)}{Pr(y_i | s_i = -1)}} = \frac{2y_i}{\sigma_{n}^{2}}
\end{equation}
Here, non-negative values of $\lambda_{i}$ indicate that a 0 was transmitted and negative values indicate that a 1 was transmitted. Further, a larger value of $|\lambda_{i}|$ indicates a higher confidence in the decision.

\subsection{Decoding scheme of SPP Code}
\begin{algorithm}\label{algo2}
    \SetKwInOut{Input}{Input}
    \SetKwInOut{Output}{Output}

    \Input{Initial LLRs $\mathbf{\Lambda}_{n}$, $\mathcal{A}$, $\mathcal{P}$, $\mathbf{w}$, $N$, $p$, $L$}
    \Output{Recovered message bits $\hat{\mathbf{d}}$}
    \hspace{2mm}\\
    $\mathcal{L} \gets \{0\}$ \tcp*[f]{A single path in the list}\\
    $PM_{0}^{(-1)} \gets 0$ \tcp*[f]{Initialize path metric}\\
    \hspace{2mm}\\
    \For{$i \gets 0$ \KwTo $(N-1)$}{
        \eIf{$i \not\in \mathcal{A}$}{
            \For{$l \gets 0$ \KwTo $(L-1)$}{
                $\hat{v}_{i}[l] \gets 0$ \\
                $\hat{u}_{i}[l] \gets \hat{v}_{i}[l]$ \\
                $\lambda_{0, i}[l] \gets$ UpdateLLR$(l, i, \Lambda[l], \beta[l])$ \\
                \If{$i\in\mathcal{P}$}{
                    $\hat{u}_{i}[l] \gets$ PrecodeOneBit$(i, \hat{\mathbf{v}}[l], \mathcal{P}, \mathbf{w}, p)$ \\
                }
                $PM_{l}^{(i)} \gets$ CalcPM$(PM_{l}^{(i-1)}, \lambda_{0,i}[l], \hat{u}_{i}[l])$\\
                $\beta[l] \gets$ UpdateBits$(\beta[l], \hat{u}_{i}[l])$\\
            }
        }
        {
            \For{$l \gets 0$ \KwTo $(L-1)$}{
                \eIf{$|\mathcal{L}| < L$}{
                    \ForEach{$l \in \mathcal{L}$}{
                        DuplicatePath$(l, i, \mathcal{P}, \mathbf{w}, p)$ \\ 
                    }
                }
                {
                    \ForEach{$l \in \mathcal{L}$}{
                        DuplicatePath$(l, i, \mathcal{P}, \mathbf{w}, p)$ \\ 
                    }
                    $\mathcal{L} \gets$ PrunePaths$(\mathcal{L})$
                }
            }
        }
    }
    $l^{*} \gets \underset{l \in \mathcal{L}}{\arg \min} \hspace{0.2cm} PM_{l}^{(N-1)}$ \\
    $\hat{\mathbf{d}} \gets$ MessageExtract$(\hat{\mathbf{v}}[l^{*}])$ \\
    \Return $\hat{\mathbf{d}}$; \\
    \hspace{2mm}\\
    \SetKwFunction{FMain}{DuplicatePath}
    \SetKwProg{Fn}{subroutine}{:}{}
    \Fn{\FMain{$l$, $i$, $\mathcal{P}, \mathbf{w}, p$}}{
        $\mathcal{L} \gets \mathcal{L} \cup \{l'\}$ \tcp*[h]{$l'$ is a copy of $l$}  \\
        $(\hat{v}_{i}[l], \hat{v}_{i}[l']) \gets (0, 1)$ \\
        $\hat{u}_{i}[l] \gets \hat{v}_{i}[l]$ \\
        $\hat{u}_{i}[l'] \gets \hat{v}_{i}[l']$ \\
        $\lambda_{0, i}[l] \gets$ updateLLR$(l, i, \Lambda[l], \beta[l])$ \\
        $\lambda_{0, i}[l'] \gets$ updateLLR$(l', i, \Lambda[l'], \beta[l'])$ \\
        \If{$i\in\mathcal{P}$}{
            $\hat{u}_{i}[l] \gets$ PrecodeOneBit$(i, \hat{\mathbf{v}}[l], \mathcal{P}, \mathbf{w}, p)$ \\
            $\hat{u}_{i}[l'] \gets$ PrecodeOneBit$(i, \hat{\mathbf{v}}[l'], \mathcal{P}, \mathbf{w}, p)$ \\
        }
        $PM_{l}^{(i)} \gets$ CalcPM$(PM_{l}^{i-1}, \lambda_{0,i}[l], \hat{u}_{i}[l])$\\
        $PM_{l'}^{(i)} \gets$ CalcPM$(PM_{l'}^{i-1}, \lambda_{0,i}[l], \hat{u}_{i}[l]')$\\
        $\beta[l] \gets$ updateBits$(\beta[l], \hat{u}_{i}[l])$\\
        $\beta[l'] \gets$ updateBits$(\beta[l'], \hat{u}_{i}[l'])$\\
    }
    \hspace{2mm}\\
    \SetKwFunction{FMain}{CalcPM}
    \SetKwProg{Fn}{subroutine}{:}{}
    \Fn{\FMain{$PM$, $\lambda$, $u$}}{
        \eIf{$u \gets \frac{1}{2}(1-sign(\lambda))$}{
            $PM \gets PM$
        }
        {
            $PM \gets PM + |\lambda|$
        }
        \Return $PM$
    }
    \caption{SCL Decoding Algorithm of SPP Code}
\end{algorithm}

The decoder of a SPP code is a combination of three entities, a LLR based successive cancellation list (SCL) decoder \cite{balatsoukas2015llr}, a \textbf{selective de-precoder} (SDP) and a message extraction unit. This is shown in Figure \ref{fig2}. 

Algorithm \ref{algo2} describes the SCL decoding algorithm for a SPP code. A SCL decoder works in multiple stages, from stage $n$ to stage $0$. $n$ refers to the input stage of the SCL decoder. $0$ refers to the output stage of the SCL decoder. Let the output of the demodulator be denoted as
\begin{equation}
    \label{eq4}
    \mathbf{\Lambda}_{n} = \{ \lambda_{n,i} \hspace{2mm}| \hspace{2mm} 0 \leq i \leq (N-1)\}
\end{equation}

The usage of $n$ in (\ref{eq4}) means that $\mathbf{\Lambda}_{n}$ is the vector of LLRs input to the SCL decoder. Further, $\lambda_{k,i}$ refers to the $i^{th}$ LLR at stage $k$ of the SCL decoder. 

SCL decoder is a sequential decoding algorithm which outputs LLRs $\lambda_{0,0}$ to $\lambda_{0,N-1}$ one by one. This is because in order to estimate the transmitted bit $\hat{u}_{i}$, SCL needs all previously decoded bits $\hat{u}_{0}$ to $\hat{u}_{i-1}$. However, in case of a SPP code, the bits $u_i$ at the input to Polar Transform are not original information or \textit{frozen} bits at each index $i$. To be specific, for each $i \in \mathcal{P}$, the bit $u_i$ is the output of precoding step as explained in Algorithm \ref{algo1}.

We now focus on the combined operation of SCL decoder and the SDP in more detail. Let $L$ denote the list size, that is, the number of parallel decoding paths of the SCL decoder. Also, let $\mathcal{L}$ be the list of decoding paths. In the beginning there is a single path in the list with index $l = 0$. The initial path metric $PM_{0}^{(-1)}$ for this path is set to 0. In the below discussion, we use the suffix $[l]$ to denote parameters for the path $l$ in the list $\mathcal{L}$.

For each path $l \in \mathcal{L}$ in the list, when the current bit index $i$ is \textit{frozen}, that is, $i \in \mathcal{F}$, the decoder knows that a 0 was transmitted. Thus, for the path $l$ in the list, $\hat{v}_{i}[l] = 0$. In addition, if $i \in \mathcal{P}$, \textbf{precoding} as per subroutine \textit{PrecodeOneBit} in Algorithm \ref{algo1} is done to generate $\hat{u}_{i}[l]$. A path metric $PM_{l}^{(i)}$ is generated by using LLR decision $\lambda_{0,i}[l]$ and $\hat{u}_{i}[l]$ as given in \cite{balatsoukas2015llr}. If the sign of $\lambda_{0,i}[l]$ matches that of $\hat{u}_{i}[l]$, then the path $l$ is not penalized. If the sign of $\lambda_{0,i}[l]$ does not match that of $\hat{u}_{i}[l]$, then the path $l$ is penalized by adding $|\lambda_{0,i}[l]|$ to the $PM_{l}^{(i)}$. In either case, $\hat{u}_{i}[l]$ is fed back to the SCL decoder to help generate $\lambda_{0,i+1}[l]$.

For each path $l \in \mathcal{L}$ in the list, when the current bit index $i$ is an information bit index, that is, $i \in \mathcal{A}$, the decoder cannot be sure which bit was transmitted. Thus, there are two options, either $\hat{v}_{i}[l] = 0$ or $\hat{v}_{i}[l] = 1$. So, the number of current paths must be duplicated. Further, all the operations mentioned above for \textit{frozen} indices are repeated for each option of $\hat{v}_i[l]$. 

Now, due to duplication, the number of paths in the list, that is, $|\mathcal{L}|$ double up. When $|\mathcal{L}| > L$, the SCL decoder selects $L$ paths with the smallest values of path metrics and discards the remaining L paths. The $\hat{u}_{i}[l]$ generated for each of the $L$ surviving paths are fed into the SCL decoder to help generate $\lambda_{0,i+1}[l]$ for each of them. 

The execution of the SCL decoder is said to end when all $N$ bits starting from the $0^{th}$ till $(N-1)^{th}$ bit have been processed. At the end, we have $L$ outputs, a vector $\hat{\mathbf{v}}[l]$ for each surviving path $l$. Let $l^{*}$ denote the path with the minimum value of path metric $PM_{l}^{(N-1)}$ among all surviving paths. The bits from this path, that is $\hat{\mathbf{v}}[l^{*}]$ are sent to the message extraction unit as shown in Figure \ref{fig2}.

Finally, the message extraction unit extracts the decoder output $\hat{\mathbf{d}}$ from $\hat{\mathbf{v}}[l^{*}]$ by picking up the bits from the information bit indices given by $\mathcal{A}$.

As one must have easily noticed, the role of SDP is critical to the decoding of SPP code under a SCL decoding paradigm. It stores the intermediate outputs $\hat{v}_{i}[l]$ for each path. It also precodes them to generate $\hat{u}_{i}[l]$ which is fed back to SCL decoder for further processing. It can also be inferred here in the interest of efficient hardware design that a single \textit{PrecodeOneBit} hardware may be used for both selective precoding and de-precoding depending upon the mode of operation for which it is configured. 

\section{Numerical results and Disccussion}

For both simulation and analysis, we consider (128, 64) codes transmitted over a BI-AWGN channel. We compare our SPP code with three contemporary polar code variants. The first code is Arikan's polar code under SCD. The second one is a CRC-Aided polar code under SCL decoding with list size $L=32$. Here, a (128, 72) polar code is combined with a (72, 64) cyclic code which acts as CRC. In other words, an 8-bit CRC is appended to the information bits before applying polar transform. The third code is Arikan's PAC code which is obtained by RM rate-profiling and a rate-1 convolutional code generated by $\mathbf{c} = [1,0,1,1,0,1,1]$. We use SCL decoding with list size $L=128$ for PAC codes as mentioned in  \cite{yao2020list} and \cite{rowshan2020polarization}.

For simulation of SPP code, we use the following parameters. The set of information bit indices $\mathcal{A}$ is designed by RM rate-profiling approach. The set of precoding indices is given by $\mathcal{P}=\mathcal{F}$. In other words, we precode the bits at \textit{frozen} indices only. The precoding vector $\mathbf{w}=[1,0,1,1,1,1,0,0,1,1,1]$. Thus, precoding length $p$ is equal to 11. Finally, we employ a SCL decoder with list size $L=128$ for decoding.

The simulation results have been shown in Figure \ref{fig1}. It can be observed clearly that both PAC code and SPP code clearly outperform CRC-Aided polar code. At a target FER of $10^{-4}$, both PAC code and SPP code are around 0.4 dB better than CRC-Aided polar code. Further, SPP code is marginally better than PAC code, but the difference in performance is only visible for FER below $10^{-4}$. At a target FER of $10^{-5}$, PAC and SPP codes are 0.25 dB and 0.23 dB away from the BI-AWGN dispersion bound approximation respectively.

It is worthwhile to note that Arikan's PAC codes can be considered to be a special case of SPP code if the code design parameters are configured as follows. The set of information bit indices $\mathcal{A}$ is designed by RM rate-profiling approach. All bit indices are precoded, that is, $\mathcal{P}=\{ 0, 1, ..., (N-1)\}$. The precoding vector $\mathbf{w}=[1,0,1,1,0,1,1]$. Thus, precoding length $p$ is equal to 7. The decoding can be done by a SCL decoder with list size $L=128$.

Although SPP codes perform only marginally better as compared to PAC codes, they have much lesser encoding and decoding complexity. This complexity reduction comes from the fact that in SPP code, precoding in the encoder and decoder happens only for bit indices indicated by $\mathcal{P}$. However, for a PAC code, such precoding happens for all $N$ indices. Especially, for a PAC code under SCL decoding, precoding at information bit indices along with path duplication leads to a lot of increase in complexity.  On the other hand, for the SPP code under SCL decoding used in our simulation, precoding does not happen during path duplication at information bit indices. This leads to a great reduction in complexity.

\section{Performance Analysis of SPP Code}
To understand the reason behind good performance of SPP code, we did a small experiment based on \cite{li2012adaptive} and repeated in \cite{yao2020list} for Arikan's PAC code. We know that the performance of a linear channel code under ML decoding is dictated by its minimum distance $d_{min}$. The higher the minimum distance, the better the performance of the code. So, we want to find out the minimum distance of SPP code and its contemporary polar code variants. 

To this end, we send all zero codewords at extremely high signal to noise ratio (SNR) and use SCL decoder to decode the channel output. For a list of size $L$, $L$ codewords will be output from the SCL decoder. Among the $L$ output codewords, one will be an \textit{all-zero} codeword, that is the transmitted codeword. The rest $(L-1)$ codewords will have as many less 1's as possible. These are the low weight codewords. As list size increases, more and more low weight codewords will emerge. 

\begin{table}[!t]
\caption{Number of low weight codewords in polar code variants}
\label{tab1}
\centering
\begin{tabular}{|l|c|c|c|c|c|c|c|}
\hline
Channel Code & $N_8$ & $N_{12}$ & $N_{16}$ & $N_{18}$ & $N_{20}$ & $N_{22}$ \\
\hline
Polar & 688 & 5376 & 193935 & 0 & 0 & 0\\
\hline
RM & 0 & 0 & 94488 & 0 & 0 & 0 \\
\hline
PAC, RM profile & 0 & 0 & 3120 & 2696 & 95828 & 238572 \\
\hline
SPP, RM profile & 0 & 0 & 2359 & 1057 & 89189 & 180966 \\
\hline
\end{tabular}
\end{table}

We keep a count of how many codewords of each weight are generated. Let the number of codewords of weight $q$ be denoted as $N_{q}$. The values of $N_{q}$ for relevant polar code variants is shown in Table \ref{tab1} for a list size $L=400000$. Due to limited computation resources, we can only compile an incomplete listing of the codeword weights. Similar to \cite{li2012adaptive}, we found that the only weights observed up to $L \leq 400000$ are those that are mentioned in Table \ref{tab1}. For example, for both PAC and SPP code, we observed only codewords with weights 16, 18, 20, 22 and 24. Due to lack of space in Table \ref{tab1}, we could not mention $N_{24} = 59784$ for PAC and $N_{24} = 126428$ for SPP respectively.

It is observed that both PAC code and SPP code should perform better than polar code because they have twice the minimum distance. Also, both PAC and SPP code should perform better than RM code. Although they have the same minimum distance 16, but both PAC and SPP code have significantly lesser number of codewords at this distance. (by a factor of 30 for PAC and 40 for SPP). Further, SPP code should perform marginally better than PAC code, because it has fewer codewords than PAC at all the distances mentioned in Table \ref{tab1}.

We can also approximate the performance of these codes under ML decoding by using \textit{truncated union bound} analysis \cite{sason2006performance}. A truncated union bound on the probability of block error can be constructed by ignoring those terms in the union bound for which the weight distribution is unknown. For a linear code of rate $R$ and a given SNR $(E_b/N_0)$, this quantity can be calculated as follows
\begin{equation}
    \label{eq5}
    P_{e}(R, E_b/N_0) = \frac{1}{\pi}\int_{0}^{\frac{\pi}{2}} \sum_{q} N_{q} exp \bigg( - \frac{RE_{b}}{N_{0}\sin^2\theta} \bigg)^{q} d\theta
\end{equation}

\begin{figure}
\centering
\includegraphics[width=3.4in]{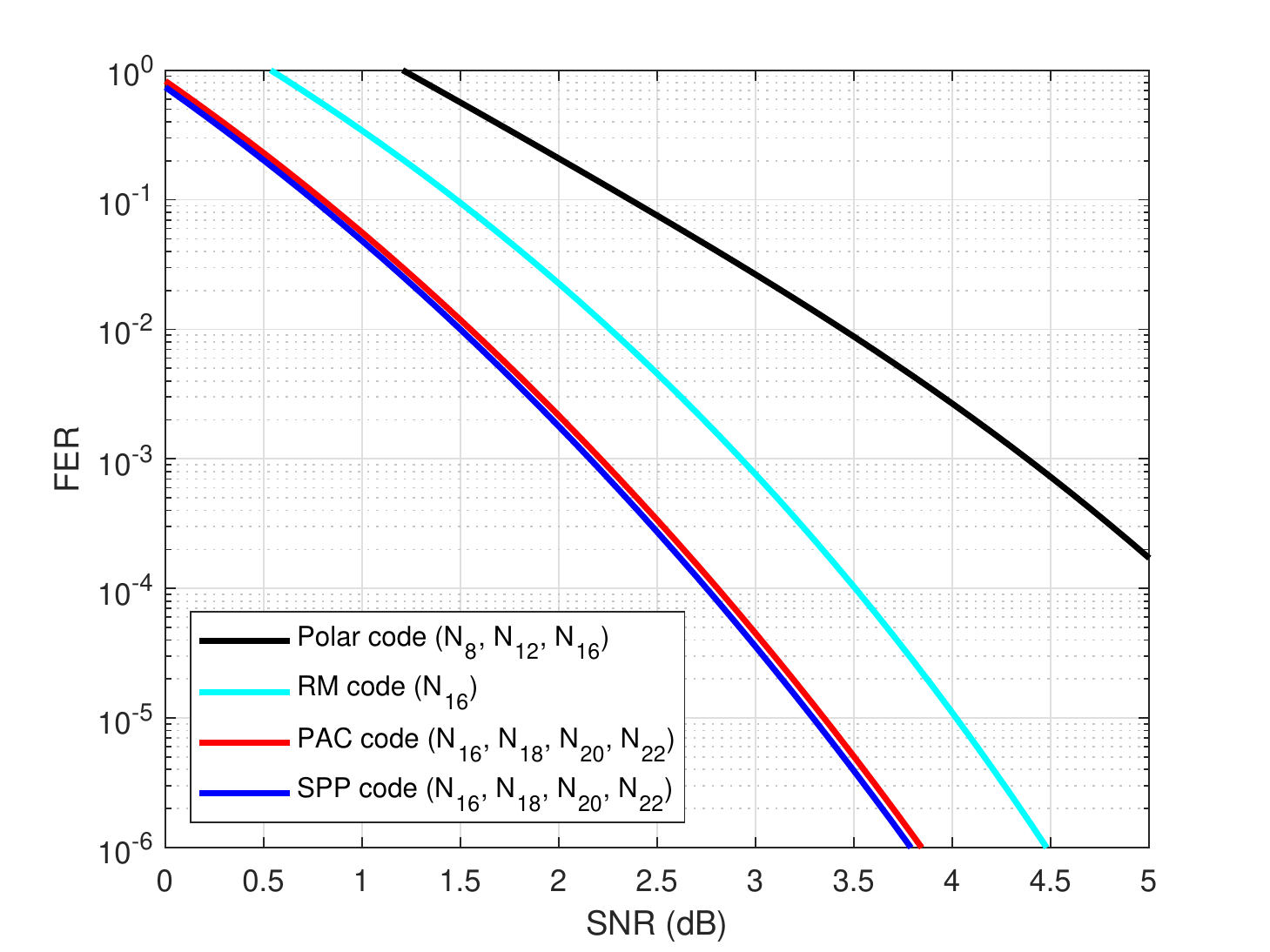}
\caption{Truncated Union Bound for (128, 64) polar code variants}
\label{fig3}
\end{figure}

Figure \ref{fig3} shows the FER performance approximation for the channel codes the incomplete weight distribution compiled in Table \ref{tab1} and applying it to (\ref{eq5}). It must be noted that both PAC and SPP code easily outperform normal polar code and RM code. Further, SPP code marginally performs better than PAC code and this is typically observed at higher SNR.

\section{Conclusion and Future Work}
In this paper, we proposed SPP code, along with its encoding and decoding schemes. In terms of FER performance for (128, 64) codes in a BI-AWGN channel, SPP code clearly outperforms other polar code variants. It also performs marginally better than Arikan's PAC codes at high SNRs. SPP code is observed to have very good distance properties similar to that of PAC code. 

There are a few open problems. First, we have only considered code of rate $R=0.5$ in this work where RM rate-profiling gives best performance. The choice of rate-profiler for other code rates needs to be explored even for a code of length 128. Second, the combination of precoding indices $\mathcal{P}$ and precoding vector $\mathbf{w}$ which can give best FER performance needs to be explored. We consider these open problems as part of future work.

\bibliographystyle{IEEEtran}
\bibliography{ref}

\end{document}